\documentclass[a4paper,12pt,useAMS]{emulateapj}
\usepackage{apjfonts,graphicx,graphics}

\newcommand{\simgt}{\lower.5ex\hbox{$\; \buildrel > \over \sim \;$}}
\newcommand{\simlt}{\lower.5ex\hbox{$\; \buildrel < \over \sim \;$}}

\shorttitle{The Y.T. Lee AMiBA}
\shortauthors{Ho et al.}

\begin{document}


\title{The Yuan-Tseh Lee Array for Microwave Background Anisotropy}


\author{
Paul T.P. Ho\altaffilmark{1,2}, 
Pablo Altamirano\altaffilmark{1}, 
Chia-Hao Chang\altaffilmark{1}, 
Shu-Hao Chang\altaffilmark{1}, 
Su-Wei Chang\altaffilmark{1}, 
Chung-Cheng Chen\altaffilmark{1}, 
Ke-Jung Chen\altaffilmark{1}, 
Ming-Tang Chen\altaffilmark{1}, 
Chih-Chiang Han\altaffilmark{1}, 
West M. Ho\altaffilmark{1}, 
Yau-De Huang\altaffilmark{1}, 
Yuh-Jing Hwang\altaffilmark{1}, 
Fabiola Iba\~nez-Romano\altaffilmark{1},
Homin Jiang\altaffilmark{1}, 
Patrick M. Koch\altaffilmark{1}, 
Derek Y. Kubo\altaffilmark{1}, 
Chao-Te Li\altaffilmark{1}, 
Jeremy Lim\altaffilmark{1},
Kai-Yang Lin\altaffilmark{1}, 
Guo-Chin Liu\altaffilmark{1,3}, 
Kwok-Yung Lo\altaffilmark{1,4}, 
Cheng-Jiun Ma\altaffilmark{1,5},
Robert N. Martin\altaffilmark{1,6}, 
Pierre Martin-Cocher\altaffilmark{1},
Sandor M. Molnar\altaffilmark{1}, 
Kin-Wang Ng\altaffilmark{1}, 
Hiroaki Nishioka\altaffilmark{1}, 
Kevin E. O'Connell\altaffilmark{1}, 
Peter Oshiro\altaffilmark{1}, 
Ferdinand Patt\altaffilmark{1}, 
Philippe Raffin\altaffilmark{1}, 
Keiichi Umetsu\altaffilmark{1,7}, 
Tashun Wei\altaffilmark{1},
Jiun-Huei Proty Wu\altaffilmark{8,7}, 
Tzi-Dar Chiueh\altaffilmark{8}, 
Tzihong Chiueh\altaffilmark{8,7}, 
Tah-Hsiung Chu\altaffilmark{8}, 
Chih-Wei Locutus Huang\altaffilmark{8,7}, 
W.Y. Pauchy Hwang\altaffilmark{8,7}, 
Yu-Wei Liao\altaffilmark{8,7}, 
Chun-Hsien Lien\altaffilmark{8}, 
Fu-Cheng Wang\altaffilmark{8,7}, 
Huei Wang\altaffilmark{8}, 
Ray-Ming Wei\altaffilmark{6}, 
Chia-Hsiang Yang\altaffilmark{8}, 
Michael Kesteven\altaffilmark{9},
Jeff Kingsley\altaffilmark{10},
Malcolm M. Sinclair\altaffilmark{9},
Warwick Wilson\altaffilmark{9}, 
Mark Birkinshaw\altaffilmark{11}, 
Haida Liang\altaffilmark{11,12}, 
Katy Lancaster\altaffilmark{11}, 
Chan-Gyung Park\altaffilmark{13}, 
Ue-Li Pen\altaffilmark{14}, \&
Jeffrey B. Peterson\altaffilmark{15}}  

\altaffiltext{1}{Academia Sinica Institute of Astronomy and
Astrophysics, P.O. Box 23-141, Taipei 10617, Taiwan} 
\altaffiltext{2}{Harvard-Smithsonian Center for Astrophysics, 60 Garden
Street, Cambridge, MA 02138, USA} 
\altaffiltext{3}{Tamkang University, Tamsui, Taipei County, Taiwan 251}
\altaffiltext{4}{National Radio Astronomy Observatory, Edgemont Road,
Charlottesville, VA 22903, USA} 
\altaffiltext{5}{IfA, University of Hawaii at Manoa, 2680 Woodlawn Dr.,
Honolulu, HI, 96822} 
\altaffiltext{6}{Composite Mirror Applications, Tucson, AZ 85710, USA}
\altaffiltext{7}{Leung center for Cosmology and Particle Astrophysics,
National Taiwan University, Taipei 10617, Taiwan} 
\altaffiltext{8}{National Taiwan University, Taipei 10617, Taiwan}
\altaffiltext{9}{Australia Telescope National Facility, Epping, NSW
Australia 1710}  
\altaffiltext{10}{Steward Observatory, University of Arizona, Tucson, AZ
85721, USA} 
\altaffiltext{11}{University of Bristol, Tyndall Avenue, Bristol BS8 1TL,
UK}
\altaffiltext{12}{Nottingham Trent University, Burton Street, Nottingham
NG1 4BU, UK}  
\altaffiltext{13}{Sejong University, Seoul, 143-747, Korea}
\altaffiltext{14}{Canadian Institute for Theoretical Astrophysics,
Toronto, ON M5S 3H8, Canada} 
\altaffiltext{15}{Carnegie-Mellon University, Pittsburgh, PA 15213 USA }

\email{pho@asiaa.sinica.edu.tw, ho@cfa.harvard.edu}


\begin{abstract}

The Yuan-Tseh Lee Array for Microwave Background Anisotropy (AMiBA) is 
the first interferometer dedicated to studying the cosmic 
microwave background (CMB) radiation at 3mm wavelength.  The choice of
3$\,$mm was made to minimize the contributions from foreground synchrotron
radiation and Galactic dust emission.  The initial configuration of 
seven 0.6$\,$m telescopes mounted on a 6-m hexapod platform was dedicated
in October 2006 on Mauna Loa, Hawaii.  Scientific operations began with
the detection of a number of clusters of galaxies via the
thermal Sunyaev-Zel'dovich effect.  We compare our data with Subaru weak
 lensing  data in order to study the structure of dark matter.  We also
 compare  our data with X-ray data in order to derive the Hubble constant.     

\end{abstract}




\section{Introduction}

The Yuan-Tseh Lee Array for Microwave Background Anisotropy
(AMiBA)\footnote{http://amiba.asiaa.sinica.edu.tw/} 
is a 
platform-mounted 7-element interferometer operating at 3-mm wavelength to
study the structure of the cosmic microwave background (CMB) radiation.
It is being constructed as part of the Cosmology and Particle Astrophysics
(CosPA) Project, funded by the Taiwan Ministry of Education Initiative
on Academic Excellence. This Excellence Initiative was aimed at
stimulating interdisciplinary research and large scale integration of
independent research programs. CosPA is designed to jump start a program
of research in cosmology, with both theory and experimental projects,
while incorporating research in high energy physics, development of
infrastructure for optical astronomy in Taiwan, as well as accessing
observing time on a 4-m class optical telescope.   

AMiBA is a collaboration between principally the Academia Sinica
Institute of Astronomy and Astrophysics (ASIAA), the National Taiwan
University (NTU) Physics and Electrical Engineering Departments, and the
Australia Telescope National Facility (ATNF).  The project was started
in 2000.  A two-element prototype was deployed in 2002 to Mauna Loa
(elevation 3396$\,$m) in Hawaii for testing of design concepts.  Site
development was completed in 2004. The AMiBA mount was delivered and
installed in 2004, while the platform was delivered and integrated in
2005.  
With the integration of the first seven elements of the array and
successful first light, the AMiBA was dedicated in October 2006, and
named after then Academia Sinica President Yuan Tseh Lee for his
important contributions in promoting the growth of astronomy in Taiwan.     
Figure 1 shows the AMiBA at the dedication ceremony. 

The aim of AMiBA is to study spatial structure in the CMB
radiation (Ho et al.~2008; Wu et al.~2008a), which carries imprints of
various physical processes in early epochs of the Universe. 
Since its initial detection by Penzias \& Wilson (1965),  
the CMB has been recognized as the definitive signature of the Big Bang
which began the expansion of the Universe.  Subsequent studies have
established the properties of this relic radiation 
after its decoupling from the matter in the early universe around
$z \simeq 1100$:
a mean temperature of $2.725\,$K (present) with minute fractional
anisotropies at the level of $10^{-5}$
(COBE, Mather et al.~1990; Smoot et al.~1992), 
and 
polarization at the level of a few to 10\% of temperature fluctuations
(DASI, Kovac et al.~2002: 
 WMAP,  Kogut et al.~2003; Page et al.~2007: Nolta et al.~2008:
 CBI, Readhead et al.~2006:
 QUaD, Pryke et al.~2008).  
In particular, the CMB structures seen on
various angular scales by COBE and then WMAP (Bennett et al.~2003;
Spergel et al.~2007; Komatsu et al.~2008) 
demonstrated that 
the angular power spectrum of CMB anisotropies is
a powerful probe of our cosmological model of the Universe.  AMiBA is
built to sample the angular range from $2'$ to $20'$, corresponding to
spherical harmonic multipoles $l=800$--$8000$, at a wavelength of 3$\,$mm,
with full polarization.  
These capabilities complement existing, on-going, and planned
experiments.
The angular scales sampled by AMiBA address
the higher-order acoustic peaks of the CMB structures to further constrain
cosmological models.  AMiBA also aims to search for, and study,
distant high-redshift clusters of galaxies whose hot intracluster gas
will distort the CMB spectrum via the thermal Sunyaev Zel'dovich effect
(hereafter SZE,
Sunyaev \& Zel'dovich 1970, 1972; Rephaeli 1995; Birkinshaw 1999;
Carlstrom, Holder, \& Reese 2002).  
The optical and X-ray surface
brightnesses of clusters of galaxies decrease rapidly with increasing
redshift due to cosmological redshift dimming, while
the detectable SZE is close to being independent of redshift because it
is a spectral distortion of the CMB radiation
which itself increases in intensity with increasing redshift as
$T_{\rm CMB}(z)\propto (1+z)$ in a standard cosmological model.
Thus SZE measurements are potentially more
sensitive than X-ray observations
for finding clusters of galaxies beyond a redshift $z\sim 1$,
and will be an important probe for 
the matter distribution in the high-redshift universe.

In this paper, we describe the design and construction of AMiBA, the
first observational results, and the scientific potential of this
instrument.        
Throughout this paper, we adopt a
concordance $\Lambda$CDM cosmology with 
$\Omega_{m}=0.3$, $\Omega_{\Lambda}=0.7$, and
$h\equiv H_0/(100\, {\rm km\, s^{-1}\, Mpc^{-1}})=0.7$.

\section{Description of the Instrument}

The basic characteristics of AMiBA are summarized in Table 1.

\subsection{Design, Construction, and Performance of AMiBA}

\subsubsection{Interferometry at 3$\,$mm}

With the funding of the CosPA/AMiBA
projects in 2000, a workshop was held to define the scientific
objectives and the design of AMiBA (Lo et al.~2001; Liang 2002).  {\it
The first design criterion was to operate at 3-mm wavelength.}  This was
to take advantage of the sweet spot at 3$\,$mm 
where the fractional SZE decrement with respect to the primary CMB is
close to its maximum (see Figure 1 of Zhang et al.~2002)
and the SZE signal is minimally-contaminated by the Galactic
synchrotron emission, dust foregrounds, and the population of cluster
and background radio sources. Operations at 3$\,$mm also complement the
wavelength coverage of other existing and planned CMB instruments:
interferometers such as CBI at 30$\,$GHz (Padin et al.~2002),
AMI at 15$\,$GHz (Scaife et al.~2008), 
SZA (Mroczkowski et al.~2008)
\footnote{http://astro.uchicago.edu/sza/} at 30 and 90$\,$GHz,
and 
VSA at 30$\,$GHz (Watson et al.~2002); 
bolometer arrays such as  
ACT,\footnote{http://www.hep.upenn.edu/act/act.html} 
APEX-SZ (Halverson et al.~2008),\footnote{http://bolo.berkeley.edu/apexsz}
and
SPT.\footnote{http://pole.uchicago.edu}

{\it
The second design criterion was to choose interferometry.}  This was a
somewhat difficult choice as many new CMB projects were then planning to
use bolometer arrays, which held the promise of a greater inherent
sensitivity because of the broad wavelength coverage and a greater speed
because of the multiple elements of the detector arrays. The choice of
interferometry was based on the desire to utilize cross correlations to
suppress systematic effects, since the ability for bolometers to
integrate down to theoretical noise was unknown at that point.
Furthermore, ASIAA had the experience of working in millimeter
wavelength interferometry from being a partner on the Submillimeter
Array (SMA) project (Ho et al.~2004). AMiBA was seen as an extension and
application of the technical capabilities within ASIAA.  Interferometry
is also a natural way to sample simultaneously the spatial structures on
various scales and to construct a map by Fourier inversion.  Imaging the
entire primary beam of the individual elements of an interferometer is 
equivalent to a multi-element detector array.  

\subsubsection{Platform Mounted Interferometer and Hexapod Drive}

{\it The third
design criterion concerns the angular sizes to be pursued.}  AMiBA
was specifically designed to sample structures in the CMB on  
small angular scales at multipoles $l=800$--$8000$, complementing in 
angular scales earlier large-sky CMB experiments (e.g., COBE, MAXIMA,
BOOMERANG) and the DASI experiment on degree/sub-degree angular 
scales ($l=140$--$900$; Leitch et al.~2002a, 2002b). 
In the left panel of Figure 2, we show the initial
compact configuration of seven 0.6$\,$m antennas ({\it small solid
circles}) on the 6$\,$m platform
({\it solid outer circle}), along with the distribution of holes ({\it
crosses})  for mounting receiver packages.
At each of the frequency channels centered at about 90 and 98$\,$GHz,
this compact configuration provides 21 simultaneous baselines 
with three baseline lengths of $d=0.61$, 1.05, and 1.21$\,$m,
corresponding to angular multipoles 
$l=2\pi \sqrt{u^2+v^2} (\equiv 2\pi d/\lambda)$ of
$l\approx 1194,2073,2394$ at $\nu_c = 94\, {\rm GHz}$. 
Also shown in the right panel of Figure 2 is the
sensitivity of the 7-element compact array
as a function of $l=2\pi
d/\lambda$ at two frequency channels, represented by
window functions (see eq. [14] of White et al.~1999)
for the three different baseline-lengths.
For the initial 
configuration of AMiBA at 3$\,$mm (see Figures 2 and 3), 
we are sensitive to the multipole range of 
$800\simlt l \simlt 3000$, which roughly overlaps with the CBI
experiment at 1cm\footnote{The achieved angular resolution of the
7-element compact AMiBA is similar to CBI, but 
the primary beam FWHM of CBI is about twice
larger than that of AMiBA.}
(Padin et  al.~2002; Mason et al.~2003; Pearson et al.~2003)
but is complementary in observing frequency.  Hence AMiBA
aimed at angular scales from $2$ to $20\,$arcmin, in order to extend the 
coverage of angular sizescales by one more order of magnitude. The
choice of 3-mm wavelength meant the required baselines were 0.6$\,$m to
6$\,$m. 
A maximum baseline of 6$\,$m immediately suggested that this interferometer
is small enough to be mounted on a platform, as pioneered by  CBI and DASI.  

{\it The fourth design
criterion was therefore to choose a carbon fiber platform for weight and
stiffness considerations.}  The platform was made with multiple holes
for mounting the receiver packages, to accommodate multiple baseline
configurations.  The layout of the receiver ports has a hexagonal
pattern in order to allow close packing, and also to utilize the
triangular patterns used by small 
interferometers such as the SMA, CBI, and DASI, in order to achieve the
most uniform $uv$-coverage.   
An ASIAA team led by Robert N. Martin and Philippe Raffin designed the
platform,
which was manufactured by Composite Mirrors Applications.  A
single rigid platform has the advantages of stable differential
pointing, a stable baseline solution without differential delay
tracking, no mutual shadowing by individual reflectors, and a single
drive system. However, a platform also means the interferometer does not
have different projected baselines because of earth rotation. 

{\it The
fifth design criterion was the choice of a hexapod mount, which provides
six degrees of freedom (which are tightly constrained) in driving the
telescope and rotating the platform.}  By rotating the reflectors with
respect to the celestial source, additional $uv$-spacings are sampled. The
rotation of the platform, described by the so-called polarization
angle of the platform, also allows us to set different orientations of
the receivers 
relative to the ground, which is useful for discriminating between
various environmental effects and checking for ground pick-up.  These
capabilities were introduced both by CBI and DASI, and we have
incorporated them.  The hexapod mount was designed and built by Vertex
Antennentechnik, Duisburg, Germany. To
protect the telescope from the elements, we use
a retractable shelter with seven steel
trusses covered with a PVC fabric, which is manufactured by American
Spaceframe Fabricator Inc.    

\subsubsection{Reflectors}

The individual parabolic reflectors are built in carbon fiber
in order to minimize their weight. The choice of 0.6m 
($0.576\,$m, to be precise) for the reflector
diameter was
driven by the desire to cover angular sizescales from $2$--$20\,$arcmin
in order to extend and overlap with existing CMB data.  The choice of
1.2$\,$m for the diameter of the second-generation reflectors
was to increase the collecting area 
and to improve the spatial dynamic range and angular resolution
for SZE observations while utilizing the entire platform.
The reflectors were designed by ASIAA, 
and manufactured by CoTech, Taichung, Taiwan.  
The design includes baffles to
shield against crosstalk between individual elements, and Gortex
covers
to shield against direct solar irradiation.  The height of the baffles
is approximately 30\% of the diameter of the reflectors so that the
secondary mirrors are well shielded.  The supports for the secondary are
attached to the baffles which are attached to the edges of the primary.
Replicating an accurate 
carbon fiber surface against a steel mold was not simple, and some hand
polishing was required.  Depositing the actual reflecting surface and a
protective coating was also not simple especially for the 1.2$\,$m size.
Nevertheless, a collaborative effort between ASIAA and CoTech was
successful in delivering the reflectors with surface accuracy better
than 50 microns. 
This is 2\% of the operating wavelength of 3$\,$mm, and would allow
efficient operation up to 1mm.   Laboratory measurements  
and outdoor beam pattern measurements
showed that the performances of the reflectors met the specifications
(for details see Koch et al.~2006).

\subsubsection{Receiver System}

{\it The sixth design criterion was to choose
heterodyne receiver systems, which operate between 86 and 102$\,$GHz.} 
This choice of frequency covers an excellent region of atmospheric
transparency.  The design, construction, and integration of the receiver
systems were performed by ASIAA staff led by Ming-Tang Chen.  Dual
polarization capabilities are provided by waveguide orthomode
transducers, which follow the circular corrugated feedhorns and the
circular-to-linear polarizers.  Each polarization is then fed into a JPL
monolithic-microwave-integrated-circuit (MMIC) InP HEMT low noise
amplifier (LNA) cooled to 15$\,$K with standard CTI22
refrigerators.  These LNAs  
performed well with measured noise temperatures of
35--50$\,$K across a 20$\,$GHz bandwidth (Weinreb et al.~1999).  
We used 
subharmonically pumped mixers (SHM) using the same MMIC technology.  
The LO/IF systems are designed and
built by Tah-Hsiung Chu and his group at the NTU EE-department.
The phase-locked LO signal at 21$\,$GHz is doubled to 42$\,$GHz and
then phase switched with Walsh functions before being combined with
the sky signal at the SHM.  Variable attenuators and amplifiers,
before and after 
mixing, control the IF levels before correlation. Slope equalizers,
phase stabilized cables, and adjustable delays are used to further
adjust the IF signals.  We measure the effective receiver temperatures
across the $2$--$18\,$GHz IF window to be 
$55$--$75\,$K.
Figure 3 shows the
AMiBA fully loaded with all the receivers and correlator modules. More
details on the receiver system are given by Chen et al.~(2008).

\subsubsection{Correlator System}

{\it The seventh design criterion was to
choose a wide band correlator with analog technology.}  This is a joint
development effort between ASIAA and ATNF, with Chao-Te Li, Derek Kubo,
and Warwick Wilson leading the effort (Li et al.~2004). A digital
correlator would have required sampling at too high a rate to be
practical.~ The AMiBA correlator uses balanced diode mixers to multiply
the signals from each pair of antennas. After application of the 4
different lags, the cross-correlated signal is then digitized and read
out with integrated circuits designed by Tzi-Dar Chiueh of the NTU EE
department.  The readout ICs demodulate the phase-switched signals and
accumulate the counts for specified times.  
The choice of a small number 
of lags is to reduce bandwidth smearing effects while maintaining the 
maximum bandwidth response through the cascade of electronics. The
bandwidth smearing effect is described by the product of the fractional
bandwidth and  
the displacement of the source from the field center.  A $10\%$ bandwidth will
keep the distortion below $20\%$ ($5\%$) at the FWHM of the primary beam of the 
$0.6\,$m ($1.2\,$m) reflectors.   
Because of the small number
of lags to cover a large bandwidth, the conversion to complex visibility
is strongly affected by gain variations over the passband, differential
delays between lags, and non-linear phase response for each lag.  The
lag-to-visibility transformation is calibrated with a noise source. This
is discussed further by Lin et al.~(2008).  


\subsubsection{Drive System and Pointing}

The hexapod drive system is more
complex than the conventional azimuth-elevation drive systems used in
radio astronomy.  The difficulty lies in the necessity of driving all
six hexapod jacks without over-extension or collisions. Vertex
Antennentechnik provided the control software.  The performance of the
drive system 
was checked via pointing and tracking tests (Koch et
al.~2008a). Pointing of the AMiBA utilized an optical telescope mounted 
on the platform.  We identified a misalignment of the anchor cone and
the mount, tilt of the optical telescope with respect to the mount,
flexure in the platform as a function of the platform polarization
angle, and local platform deformation.  We also measured the
repeatability of pointing on both short and long timescales, and we
performed photogrammetry to measure the stability of the platform.
Pointing was found to be repeatable at the $4\arcsec$ level on a timescale of
several hours.  The absolute rms pointing error appears to be about
$0.8\arcmin$ for the platform set at zero polarization angle, and up to
3$\arcmin$ 
if the platform is allowed to rotate to different polarization angles.
However, with the implementation of an interpolation table, absolute rms
pointing can be reduced to $0.4'$ over all sky.  This is less than $10\%$
of the primary beams of the 0.6$\,$m reflectors, and is within
specifications for the operation of the 7-element compact array
(Koch et al.~2008a).  However, when the 1.2$\,$m reflectors
are deployed, pointing needs to be improved by a factor of 2.

\subsubsection{Deformation of the Platform}

The segmented approach instead of a
monolithic design for our carbon fiber platform resulted in unforeseen
difficulties.  
The bolted joints between the six outer segments were not stiff enough.
In spite of efforts to strengthen the joints with
additional plates and brackets, the platform still deforms under
operational loads to a level beyond our specifications.  
A saddle-shaped deformation pattern is present with an amplitude
dependent on the hexapod azimuth, elevation and polarization position. 
Fortunately, the deformation appears repeatable when measured with
photogrammetry which means that it can be modeled.  Maximum deformation
at the edge of the platform under simulated full loading is 0.38$\,$mm.  The
$\,$maximum deformation in the inner 3$\,$m of the platform is more modest at
0.120$\,$mm, and within specifications for the
operation of the 7-element compact array.
The deformation is due principally to the platform not being
stiff enough, and also because the stresses from the drive system are
transmitted to the platform in spite of a steel interface ring (Koch et
al.~2008a).   The immediate ramification is that the individual primary
beams of the interferometer elements will be mis-pointed relative to
each other.  This is equivalent to the primary mirror of a single dish
telescope being deformed under gravity or atmospheric distortion.  The
equivalent adaptive optics approach for an interferometer involves
adjusting the gains and phases of the individual elements either through
a look-up table or via self calibration if the signals are strong
enough.   
The advantage of an interferometer is that we can adjust the
signals during the postprocessing phase before adding or multiplying the
signals from different baselines. 
For our current configuration, such 
correction schemes for the platform deformation do not need to be implemented.  

\subsubsection{System Performance}

Calibration of the AMiBA system was led by Kai-Yang Lin and Chao-Te Li.
The receiver temperatures were measured by the standard hot/cold load
method to be 55 to $75\,$K. The contribution from the sky and ground
pickup
to the total system temperatures is about $25\,$K.  Since the correlator
only has 4 lags, only 2 frequency channels can be extracted. The
response of the correlator is therefore not easy to measure, and we use
planets to calibrate the correlator response periodically. We find the
complex gain to be stable to 5\% in gain and 0.1 rad in phase on time
scales of a few hours.     

Using the detected fluxes for Saturn and Jupiter, the overall system
efficiency was estimated to be $0.3$--$0.4$ for each of the baselines of the
interferometer.  This efficiency accounts for all losses due to
illumination, blockage, spillover, alignment of the reflectors with
respect to the radio axis of the platform, deformation of the platform,
and pointing, as well as the narrower correlator response (Lin et
al.~2008).  The losses in front of the receiver account for the
significant  contributions from the sky and ground pickup to the system
temperatures.   
In two-patch differencing observations
a sensitivity of about $63\,$mJy is achieved in 1$\,$hour of on-source 
integration under good sky conditions (Chen et al.~2008; Lin et al.~2008).

\subsubsection{Gaussianity of the Noise}

Since the SZE signals and the CMB anisotropy are quite faint, the
behavior of the noise is very important.  This analysis has been led by
Hiroaki Nishioka.  We conducted tests of blank sky data as well as data
where the inputs are terminated by absorbers. Statistical analysis of
different samples of such data showed no significant differences.  Time
variable signals from the correlator due to electronics and ground
pickups can be seen in the various datasets.  However, a power spectrum
analysis demonstrated white noise behavior for frequencies between
10$^{-4}$ and $1\,$Hz.  Hence two-patch differencing of the sky signal at
600~sec intervals is adequate to remove the slowly varying
contaminations.   

By cross correlating the outputs from the different lags of the
correlator, we find only a weak correlation at less than 10\%.  By
dividing the datasets into smaller samples, we verify that the sample
variance is consistent with Gaussian random noise which would integrate
down over time.  We also applied the Kolmogorov-Smirnov (K-S) test to
our cluster data, and we find that 90\% of the data are
consistent with Gaussian behavior at the 5\% significance level.  The
data which fail the test at 5\% level are often found to be associated
with hardware problems, and we use this criterion to discard
questionable samples.  More details on the Gaussianity tests are given
in Nishioka et al.~(2008).

\subsubsection{Contamination on the Observed SZE Signals}

Since the SZE signals are faint, they can be confused with the CMB
anisotropies and foreground emission structures.  The analysis of
possible contaminations has been led by Guo-Chin Liu. Examining the WMAP
data, we can see that CMB structures on the scale of a degree are
suppressed by our two-patch observing technique, while contamination
from Galactic emission is typically fainter than the CMB emission by an
order of magnitude.  From WMAP data, we also estimate possible
contamination from primary CMB anisotropies, by analyzing 500 simulated
CMB fields, for the sizescales which we sample.  We find contamination
at the level of $13$ to $90\,$mJy for the sampled multipoles 
of $l=1200$ to $2400$ (at $94\,$GHz).

We also estimated potential contamination from foreground point
sources.  The best way to correct for this issue is to have high
angular
resolution interferometer data at the observed frequencies.  Lacking
such data, we extrapolate the contributions from known point sources in
low frequency radio surveys.  Since spectral indices are not known
accurately to $3\,$mm, a statistical approach is used to estimate the
contamination in each cluster from known point sources.  
In all clusters, a net positive contribution of the point sources was found
in our main-trail/lead differencing AMiBA observations,
indicating that there are more radio sources
towards clusters than in the background (as expected; see Coble et al. 2007).
The corrections for
contamination are described more fully in Liu et al.~(2008).

\subsection{First Science Results from the AMiBA}

\subsubsection{First Light}

First light with the 7-element AMiBA was achieved in September 2006. The
array of 0.6$\,$m reflectors was in the close-packed configuration on the
platform (Figures 2 and 3).  
The image of Jupiter shown in Figure 4 is the first
end-to-end test of the system based on drift-scan data, 
including the pipeline for data analysis.
The image was constructed from drift scans at four
equally-divided platform polarization angles 
to provide better $uv$-coverage.  
Only the transit data of about 0.23$\,$s exposure time 
from each drift scan were used, 
yielding an overall noise level in the dirty map of about 3$\,$Jy,
much lower than the Jupiter flux density of 844$\,$Jy.
Successful images were obtained for Saturn, Venus,
and the Crab Nebula, all with fluxes on the order of 200$\,$Jy. In
particular, the measured angular size of the Crab Nebula is consistent
with existing optical and radio images.  In December 2006, using Saturn
as the calibrator, Uranus was imaged with a flux
density of $11 \pm 4\,$Jy (at $2\sigma$ confidence level; in 16 $1\,$s
exposures), consistent with the expected level of $7.3\,$Jy.    

\subsubsection{Imaging Clusters of Galaxies}

At the achieved sensitivity, the close-packed 7-element configuration of
AMiBA is suitable for imaging galaxy clusters via the SZE.  During 2007,
we imaged 6 clusters, chosen to be relatively nearby (a median redshift
of $z\sim 0.2$), already
detected in SZE, and resolvable with AMiBA.  Target selection and data
analysis are described in Wu et al.~(2008b).    
To detect the much fainter SZE signals from clusters of
galaxies, we need to integrate longer.  
At this level of sensitivity, 
emission from local terrestrial sources such as buildings and the ground
will enter through the system either through the sidelobes of the
antenna response 
or via scattering into the primary beam. 
The interferometer has the
advantage over a bolometer array in that the cross correlation will
suppress extended low level emission which enters the system through the
sidelobe response of the reflectors.  However, structures on angular
scales corresponding to the interferometer baselines will still persist.
The design of the baffles around the edges of the reflectors helps
further suppression, but we did not exercise the option of installing
ground shields.  Instead we employ a two- or three-patch observing
technique where the target source is preceded and/or followed by
tracking over the same range of azimuth and elevation just traversed by
the source.  This allows us to subtract and cancel the terrestrial
ground and sky emission.  This is the AMiBA equivalent of the
position-switching technique used in single-dish radio astronomy,
where 
we track specifically the same part of the atmosphere. 
 As
shown in Figure 5, and in Wu et al.~(2008b), this procedure is quite
successful and we have mapped the SZE decrement towards a number of
clusters.  These detections serve to demonstrate the scientific
potentials of AMiBA. 

\subsubsection{Distribution of Mass and Hot Baryons in Cluster Environment}

AMiBA will be sensitive to structures as large as $20\arcmin$
with resolution as
high as $2\arcmin$ 
so that SZE structures can be resolved.  Even at the current
angular resolution of $6\arcmin$ FWHM, 
an extended, elliptical shape can be seen for some of
the clusters.  These shapes can be compared to the results from weak
gravitational lensing or X-ray imaging. The SZE and X-ray images are
sensitive to the distribution of the hot intracluster gas, while the weak
lensing images probe the distribution of all the gravitational mass
including the dominant ``invisible'' dark matter.  
Furthermore, the X-ray emission is sensitive
to the square of the electron density while the SZE measures the
thermal electron pressure projected along the line-of-sight, 
linearly tracing the electron column density.
Hence, the complementary X-ray, the SZE, and the weak
lensing data, probe progressively further out in the mass distribution.

For four of the massive clusters in our sample (A1689, A2142, A2261, A2390), 
for which
high-quality Subaru weak lensing data are available,
we have compared the SZE
results with the Subaru weak lensing measurements  (Umetsu et al.~2008). In
the case of Abell cluster A2390, the elliptical shape seen in SZE is
consistent with the shape of the dark matter distribution as deduced
from weak lensing data.  

For this comparison, 
A2142
is of particular interest since this is our brightest, most-nearby
(resolvable)  
SZE cluster at $z=0.091$, known as a merging cluster with two X-ray {\it cold
fronts},  which are sharp edges in X-ray dense cores (Markevitch et
al.~2000). In Figure 6 we compare for A2142
our AMiBA SZE map with the projected
mass distribution ({\it white contours}) 
as deduced from our weak lensing analysis. 
Here the mass map is smoothed to a resolution of $2\arcmin$ FWHM.
Our AMiBA SZE map shows an elliptical structure extending in the
northwest (NW) - southeast (SW) direction, 
similar to X-ray and weak lensing distribution shapes (Figure 6; see
also Figure 10 of Okabe \& Umetsu 2008). The
almost $20\arcmin$ angular extent of the cluster SZE
justifies the use of observing with antennas as small as $0.6\,$m, so
that the observations are sensitive to large scale structures.
Further, relative positions (between pixels)
are important in such maps of extended
structures, where the interferometer has the inherent advantage of
relying on the phase to determine relative positions, while single dish
studies might be affected more strongly by pointing of the telescopes. 

In the cluster A2142, a mass subclump located $\sim 10\arcmin$ 
($\sim 710\,$kpc$h^{-1}$ at the cluster redshift)
to the NW of the cluster center has been detected from the Subaru
weak lensing analysis of Okabe \& Umetsu (2008) and Umetsu et
al.~(2008). This NW mass subclump,  
lying $\sim 5\arcmin$ ahead of the NW- edge of the X-ray dense
core, has no X-ray counterpart in Chandra and XMM-Newton observations,
but is associated with a slight excess
of cluster sequence galaxies
(see Figure 10 of Okabe \& Umetsu 2008).
Our AMiBA map exhibits slight excess SZE signals in the NW region (Figure 6)
at $2-2.5\sigma$ significance levels. We note this slight excess SZE 
appears extended over a couple of synthesized beams, although the
per-beam significance level is marginal.
Good consistency found between the SZE and weak lensing
maps is encouraging, and may suggest that the NW excess SZE is a
pressure increase in the ICM associated with the moving NW substructure.
Clearly further improvements in both sensitivity and
resolution of the SZE mapping (see \S\ref{subsec:expansion})
are needed to better constrain
the merger geometry and physical properties of the merging
substructure.  
 This however demonstrates the potential of SZE observations as a powerful
tool for measuring the distribution of ICM in cluster outskirts
where the X-ray emission measure ($\propto n_e^2$) is less sensitive.
This also demonstrates the potential and reliability of AMiBA, and 
the power of multiwavelength cluster analysis 
for probing the distribution of mass and baryons in
clusters. 


Finally, for our small sample with a mean virial mass of
$\langle M_{\rm
vir}\rangle=(1.2\pm 0.1)\times 10^{15}M_{\odot} h^{-1}$, 
the hot gas mass 
fraction is about $(13\pm 3)\%$ (Umetsu et al.~2008). 
As compared to the cosmic baryon fraction $f_b=\Omega_b/\Omega_m$ of
 $\simeq 17\%$,
as deduced from the WMAP
5-year data (Dunkley et al.~2008), 
possibly $(22 \pm 16)\%$ of the baryons are missing from the hot
cluster environment (Umetsu et al.~2008).  
This missing cluster baryon fraction is partially made up by
observed stellar and cold gas fractions of
$\sim$ several $\%$ in our X-ray temperature range ($T_X>8\,$keV).

\subsubsection{Estimating the Hubble Constant}

By comparing the SZE and X-ray imaging results, we can deduce a value
for the Hubble constant $H_0$.  Since the SZE is induced by inverse-Compton
scattering of the CMB photons by hot electrons in the cluster
environment, it is sensitive to the total column of electrons.  The
X-ray emission is due to the Bremstrahlung process between protons and
electrons, and is therefore sensitive to the square of the electron
density. 
By combining these two relations, the angular diameter distance can be 
deduced.    
However this derivation, and the subsequent measurement of $H_0$ depends
on the assumed geometry of the emission region.  We have derived a value
of $H_0$ from our sample of SZE results combined with X-ray data.  The
uncertainty is high because of the small size of our sample, 
and the
bias from the assumed geometry.  However, the deduced value is
consistent with other SZE constraints, 
and is an indicator of integrity of the AMiBA data so far.
Better limits will come with a much larger sample to be obtained in the
future.   
 
\subsection{Expansion to 13-elements}
\label{subsec:expansion}

While the initial 7-element $0.6\,$m reflectors have been commissioned, we
are proceeding with the expansion of the AMiBA to its 13-element
configuration.  In this configuration, we will upgrade from $0.6\,$m to
$1.2\,$m antennas.  This will increase the collecting area by a factor of
$\sim 7.4$, and the speed of the interferometer by a factor of almost 60
in single pointed observations.  We
will place the 13 elements over the platform to generate the longest
possible baselines, which will result in angular resolutions up to $2\arcmin$.
The correlator is also being expanded in order to handle the larger
number of cross correlations.  

In this second phase of AMiBA operations, the first science target will
be to measure the angular power spectrum of CMB temperature anisotropies
to the higher multipole numbers in order to examine the shape at and
beyond the third acoustic peak ($l\sim 800$; Nolta et al.~2008), where
data taken with the 7-element close-pack 0.6$\,$m configuration will be
combined with upcoming 13-element 1.2$\,$m data.
Accurate measurements of the angular power spectrum of
CMB temperature anisotropies (Park et al.~2003)
through $l\sim 4000$ up to $l\sim 8000$ 
will allow us to see secondary effects such as the SZE (Lin et al.~2004) 
and possible cosmic string structures (Wu 2004).  
The second science target will be to resolve cluster SZE
structures on the sky in order to compare with dark matter structures as
deduced from weak gravitational lensing studies 
(Umetsu \& Futamase 2000;
Broadhurst et al.~2005, 2008; 
Umetsu \& Broadhurst 2008;
Okabe \& Umetsu 2008).  
The third science target will be to survey for the
distribution of galaxy clusters via the SZE (Zhang et al.~2002;
Umetsu et al.~2004). 
To obtain redshifts of cluster candidates, optical follow up
observations will be conducted with ground-based telescopes.   

At the time of the publication of this paper, the expansion is already
in progress.  We anticipate first operations during 2009.


\section{Conclusion}

In this introductory paper, we have presented the design and
construction of the Yuan-Tseh Lee AMiBA project.  The telescope is now
in scientific operation, and the first science results are the detection
and imaging of six galaxy clusters via the Sunyaev Zel'dovich effect.    
Our companion papers elaborate on various aspects of our 
results.  Chen et al.~(2008) describe the technical design
of the instruments in detail.  Wu et al.~(2008b) describe the
observations and data analysis for the first year of AMiBA SZE data.  
Lin et al.~(2008) examine the system performance, and discuss the
astrophysical properties of planetary calibrators. 
Koch et al.~(2008a)
describe the performance of the hexapod mount and the carbon fiber
platform. Nishioka et al.~(2008) examine the integrity of the AMiBA data
and its statistical properties.  Liu et al.~(2008) address the issues of
foreground and primary-CMB contamination in AMiBA SZE observations.
Umetsu et al.~(2008) 
discuss the combination of AMiBA SZE and Subaru weak lensing
data in order to probe the structure of dark matter and to derive the
cluster gas mass fractions.
Koch et al.~(2008b) derive the value of the
Hubble constant $H_0$ from AMiBA and X-ray data on the clusters.  
Huang et al.~(2008) derive cluster scaling relations between the SZE and
X-ray observables.
Finally, Molnar et al.~(2008) discuss the potential of AMiBA data in its
13-element configuration to constrain the intra-cluster gas distributions.


\acknowledgments
We thank the Ministry of Education, the National
Science Council, and the Academia Sinica for their support of this
project.  We thank the Smithsonian Astrophysical Observatory for hosting
the AMiBA project staff at the SMA Hilo Base Facility.  We thank the
NOAA for locating the AMiBA project on their site on Mauna Loa.  We
thank the Hawaiian people for allowing astronomers to work on their
mountains in order to study the Universe.

 \clearpage

 \begin{deluxetable}{ll}
 \tabletypesize{\scriptsize}
 \tablecolumns{2}
 \tablecaption{
  \label{tab:amiba}
  Basic Characteristics of the AMiBA
 } 
 \tablewidth{0pt} 
 \tablehead{ 
  \colhead{Components} &
  \colhead{Specifications}
 }
 \startdata 
 Interferometer elements         & 7 0.6-meter (13 1.2-meter), 
				   f/2.0  Cassegrain\\
 Telescope mount                 & hexapod \\
 Telescope backup structure      & 6-meter carbon fiber platform \\
 Primary reflector               & monolithic carbon fiber\\
 Surface accuracy                & 30 microns rms \\
 Secondary reflector             & carbon fiber, fixed\\
 Array configuration             & rings at 0.6$\,$m spacings\\
 Available baselines             & 0.6 - 6.0 meters\\
 Operating frequencies           & 94$\,$GHz\\
 Maximum angular resolution      & $2'$\\
 Primary beam field of view      & 23$'$ (11$'$)\\
 Receiver band                   & 86--102$\,$GHz\\
 Number of receivers             & 7 (13) dual polarization MMIC IP HEMT\\
 Correlator                      & 4-lag analog 2 $\times$ 21 (78) baselines \\
 Point source sensitivity        & 63 (8) mJy in 1 hour on-source integration\\
 \enddata
 \tablecomments{The values in parenthesis will apply after the system upgrade
  to 13-elements with 1.2-meter antennas. The point source sensitivity
  here is for two-patch differencing observations.}
 \end{deluxetable}

 \clearpage

 \begin{figure}
 \begin{center}
   \includegraphics[width=140mm,angle=0]{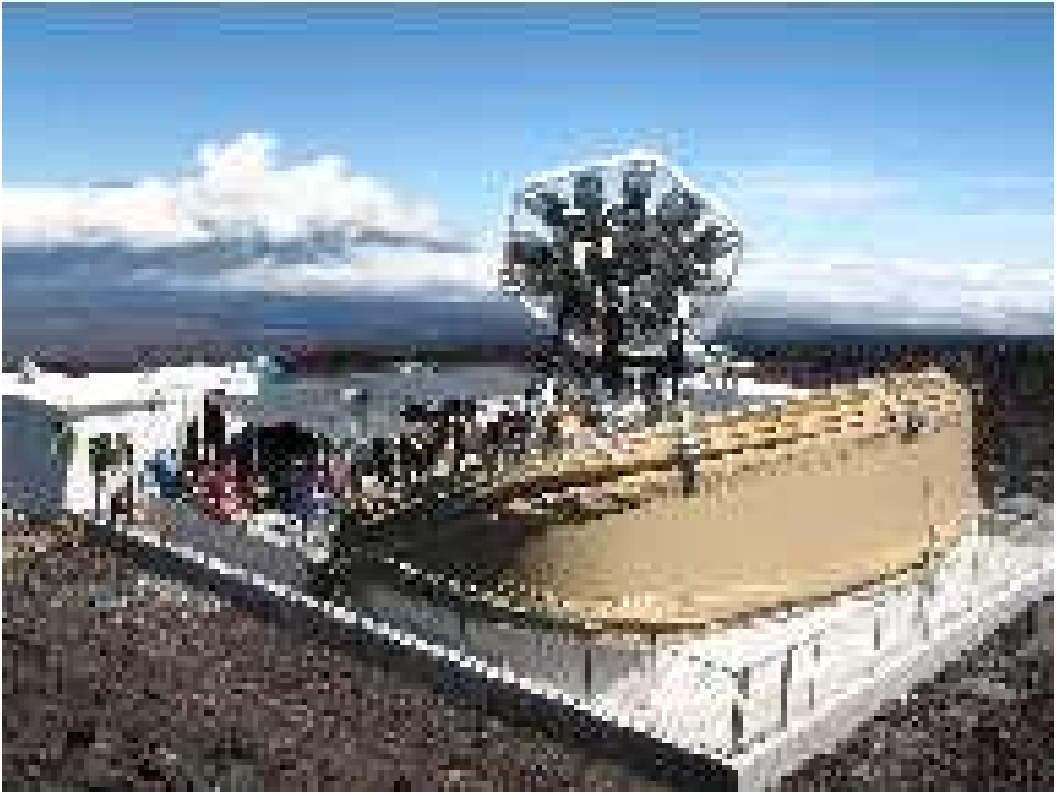}
  \end{center}
 \caption{View of the AMiBA telescope on Mauna Loa, in October 2006,
  during dedication.     
 \label{fig1}}
 \end{figure}

 \begin{figure}
 \begin{center}
   \includegraphics[width=140mm,angle=0]{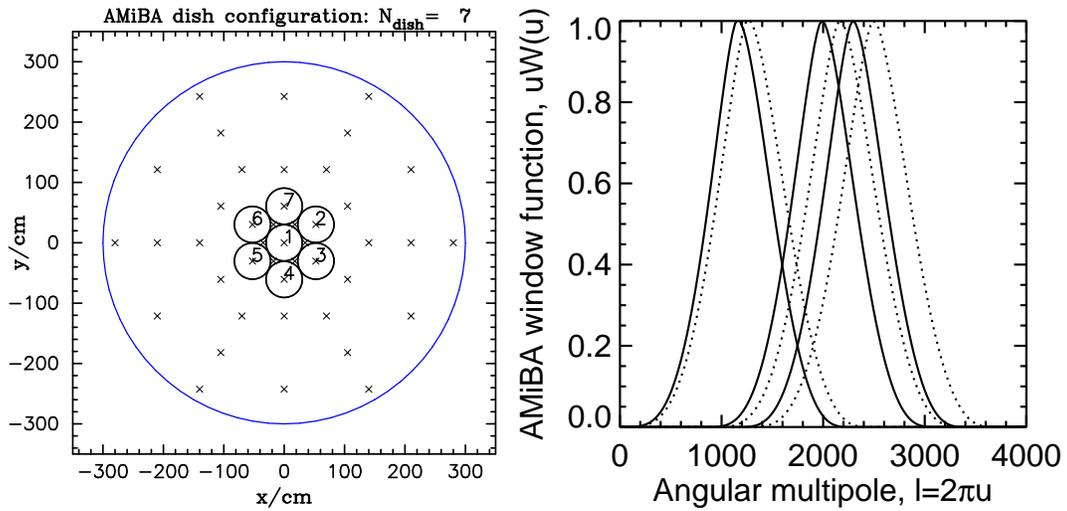}
  \end{center}
 \caption{
 {\it Left}: 
 Initial compact configuration of 
 seven 0.6$\,$m antennas 
 ({\it small solid circles}) on a 6$\,$m single platform ({\it outer solid
  circle}). {\it Right}: Sensitivity of the 7-element AMiBA as a
  function of spherical harmonic multipole $l=2\pi d/\lambda$ at two
  frequency channels $90$ ({\it solid}) and 98 ({\it dashed}) GHz,
  shown by window functions for three baselines with different lengths
  (see eq. [14] of White et al.~1999).
 \label{fig2}}
 \end{figure}

 \begin{figure}
 \begin{center}
   \includegraphics[width=70mm,angle=0]{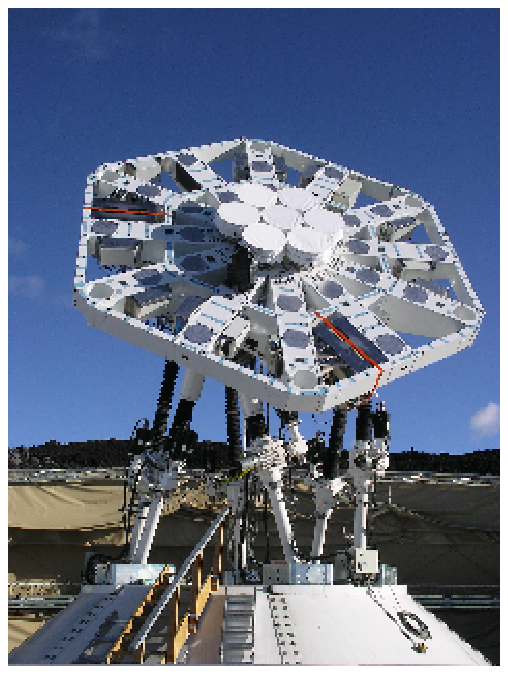}
   \includegraphics[width=70mm,angle=0]{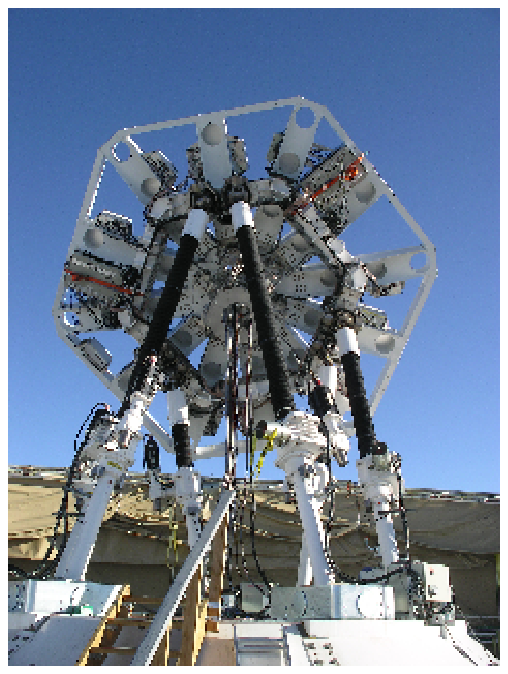}
  \end{center} 
 \caption{
 A close-up of the AMiBA telescope.
 The left panel shows the initial configuration of seven 0.6$\,$m
  antennas  co-mounted on a 6$\,$m platform.
 Shown in the right panel are
 the receiver
  packages mounted on the platform together with various electronics such
  as the correlator and LO/IF systems. The reflectors and receivers can
  be deployed at various locations on the platform in order to achieve
  different projected baselines.   
 \label{fig3}}
 \end{figure}

 \begin{figure}
 \begin{center}
   \includegraphics[width=120mm,angle=0]{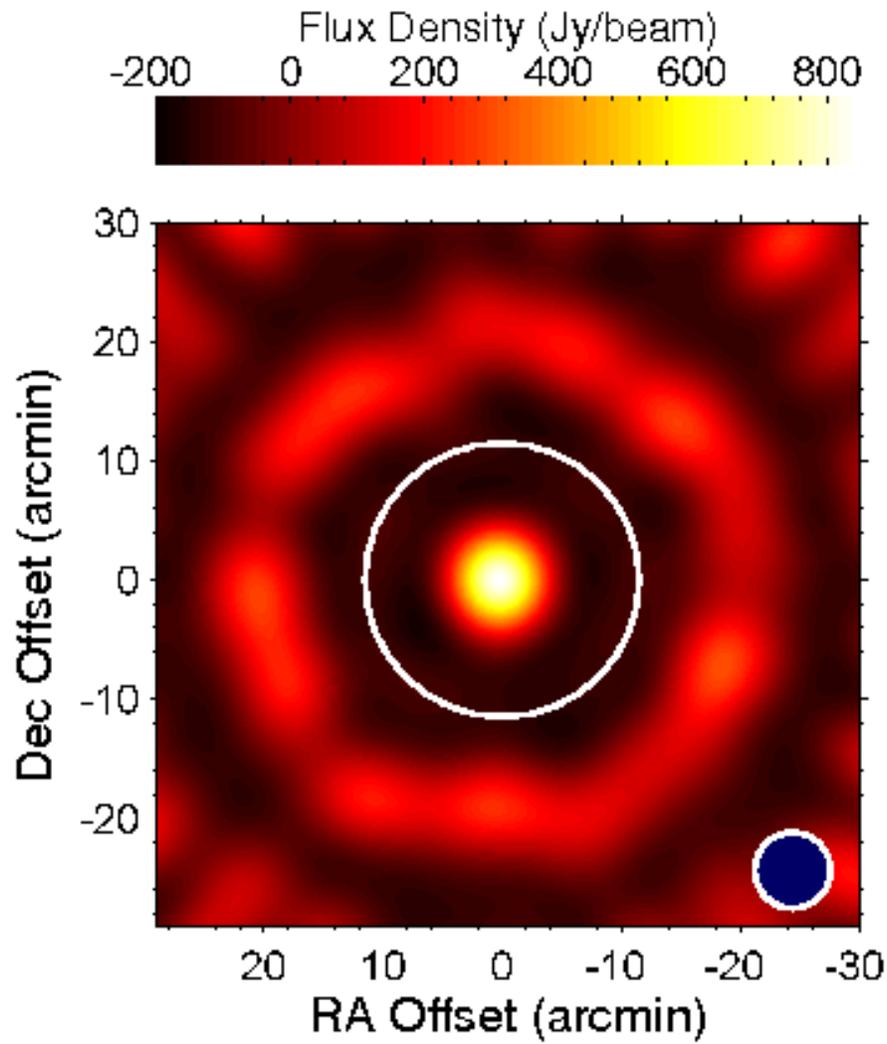}
  \end{center}
 \caption{The first light image of Jupiter obtained by AMiBA in September
  2006.  This was a verification of the receiver and correlator systems
  as well as the pipeline software developed to calibrate and image the
  interferometer data.  This is an un-cleaned, ``dirty'' image.
  The white circle indicates the field of view,
  while the blue region at the bottom-right corner shows
  the FWHM of the synthesized beam, which is about $6'$.
 \label{fig4}}
 \end{figure}

 \begin{figure}
 \begin{center}
   \includegraphics[width=150mm,angle=0]{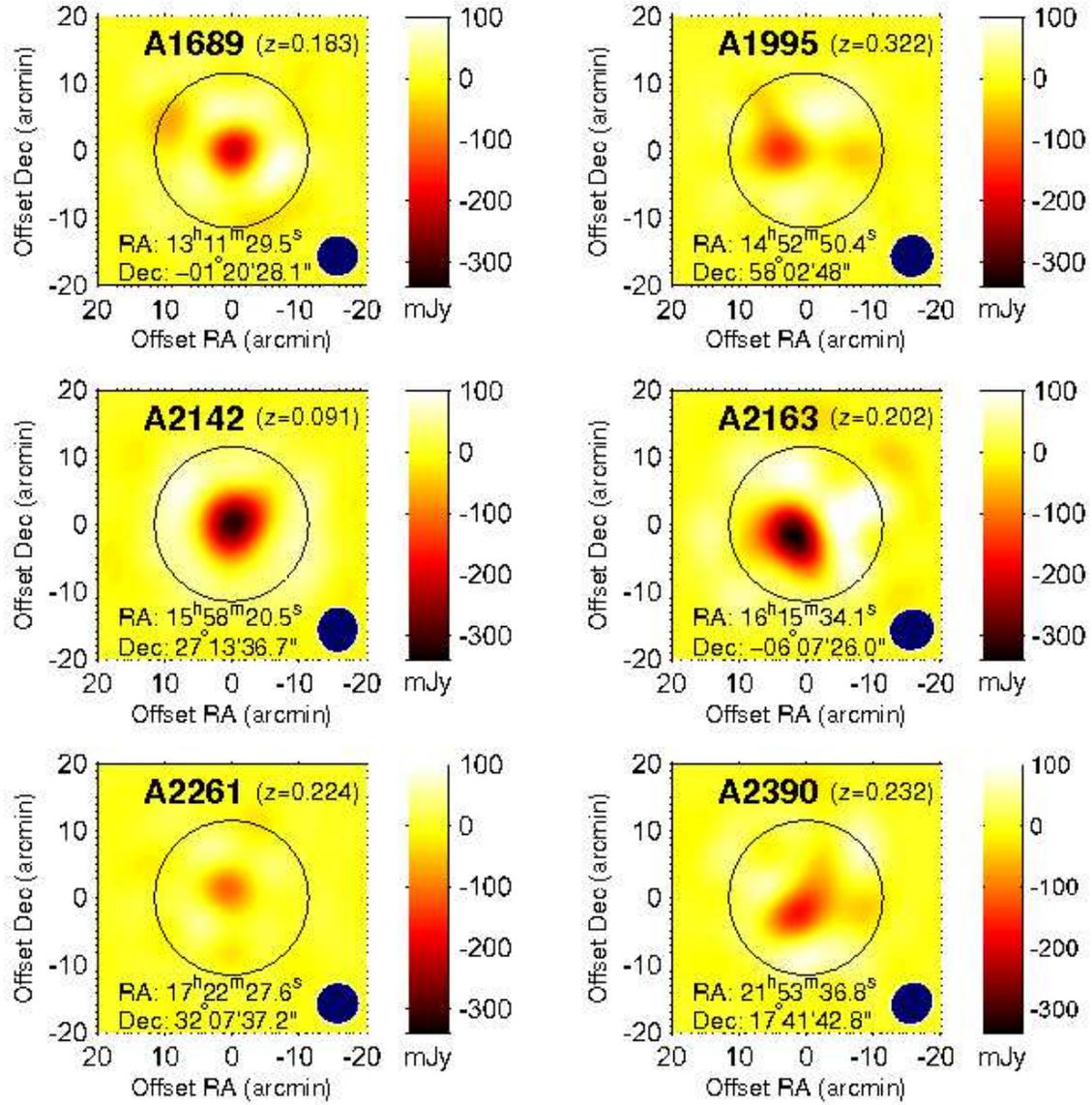}
  \end{center}
 \caption{
 The first AMiBA images of the SZE decrement towards six massive
  clusters of galaxies, A1689, A1995, A2142, A2261, and A2390. 
 \label{fig5}}
 \end{figure}

 \begin{figure}
 \begin{center}
   \includegraphics[width=130mm,angle=0]{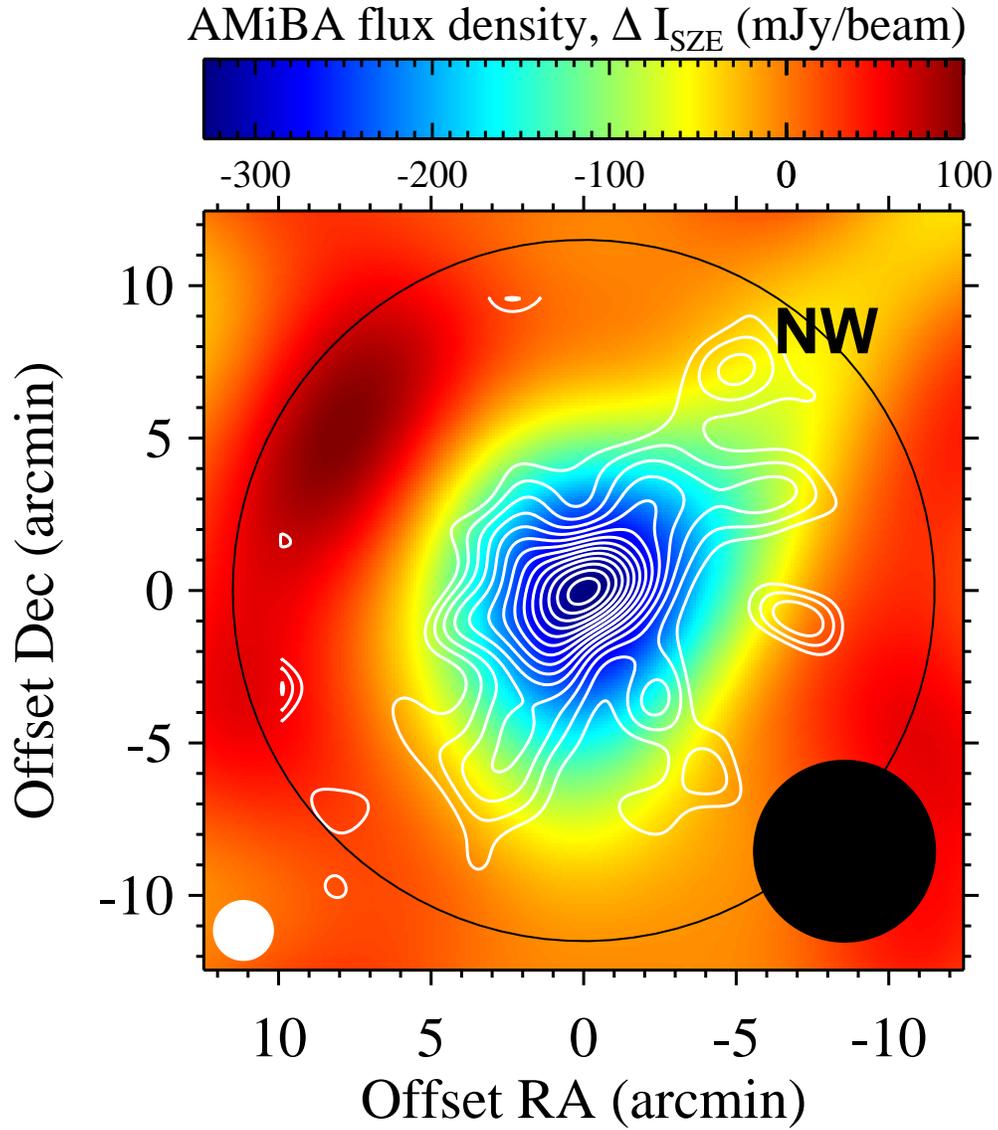}
  \end{center}
 \caption{
The cleaned AMiBA map of the cluster A2142 ($z=0.091$), revealing a
  strong SZE decrement of about $-330\,{\rm mJy}\,{\rm beam}^{-1}$ in
  the cluster center (Wu et al.~2008b).
The field size shown is $25\arcmin$, corresponding to $\simeq
  1.8\,$Mpc$h^{-1}$ at the cluster redshift.
The black circle indicates the size of the 
AMiBA field-of-view ($23\arcmin$ FWHM), and the black filled circle at
  the bottom-right corner shows the size of the AMiBA synthesized beam
  ($6\arcmin$ FWHM).
The residual rms noise level in the cleaned map is $\simeq 23\,$mJy (Wu
  et al.~2008b).
Overlaid are the contours ({\it white}) of the projected mass
  distribution reconstructed from Subaru weak lensing data
(see Umetsu et al.~2008; Okabe \& Umetsu 2008).
The contours are
spaced from the $2\sigma$ noise level at intervals of $1\sigma$. 
Shown at the bottom-left corner is 
the $2\arcmin$ FWHM of the Gaussian smoothing kernel ({\it white filled
  circle}) used for mass reconstruction. 
The weak lensing map shows a mass subclump in the northwest region
  located about $10\arcmin$ northwest from the cluster center. 
A slight excess of the SZE signal, extending in the northwest direction,
 is seen in the northwest region at
  $2-2.5\sigma$ levels.
 \label{fig6}}
 \end{figure}

\end{document}